\documentclass[prd,reprint,showpacs,showkeys]{revtex4}
\usepackage{graphics}
\usepackage{eurosym}
\usepackage{amsfonts}
\usepackage{amssymb}
\usepackage{amsmath}
\usepackage{graphicx}
\usepackage[font={footnotesize,it}]{caption}
\usepackage[colorlinks=false,
            linkcolor=red,
            urlcolor=red,
            citecolor=blue]{hyperref}

\usepackage{color}

\begin{document}

\title{Role of pressure anisotropy on relativistic compact stars}

\author{S. K. Maurya}
\email{sunil@unizwa.edu.om}
\affiliation{Department of Mathematical and Physical Sciences,
College of Arts and Science, University of Nizwa, Nizwa, Sultanate
of Oman}

\author{ Ayan Banerjee}
\email{ayan_7575@yahoo.co.in}
\affiliation{Department of Mathematics, Jadavpur University, Kolkata 700032, West Bengal, India}
\affiliation{Astrophysics and Cosmology Research Unit, University of KwaZulu Natal, Private Bag X54001, Durban 4000,
South Africa.}

\author{ Sudan Hansraj}
\email{hansrajs@ukzn.ac.za}
\affiliation{Astrophysics and Cosmology Research Unit, University of KwaZulu Natal, Private Bag X54001, Durban 4000,
South Africa.}

\date{\today }

\begin{abstract}
 We investigate a compact spherically symmetric relativistic body with anisotropic particle pressure profiles.  The distribution possesses  characteristics relevant to modeling compact stars within the framework of general relativity. For this purpose, we consider a spatial metric potential of  Korkina and Orlyanskii [Ukr. Phys. J. 36, 885 (1991)] type in order to solve the Einstein field equations. An additional prescription we make is that the pressure anisotropy parameter takes the functional form proposed by Lake [Phys. Rev. D 67, 104015 (2003)]. Specifying these two geometric quantities allows for further analysis to be carried out in determining unknown constants and obtaining a limit of the mass-radius diagram, which adequately describes compact strange star candidates like Her X-1 and SMC X-1. Using the anisotropic Tolman-Oppenheimer-Volkoff equations, we explore the hydrostatic equilibrium and the stability of such compact objects. Then, we investigate other physical features of this models, such as the energy conditions, speeds of sound and compactness of the star in detail and show that our results satisfy all the required elementary conditions  for a physically acceptable stellar model. The results obtained  are useful in analyzing the  stability of other anisotropic compact objects like white dwarfs, neutron stars, and gravastars.
\end{abstract}


\maketitle

\section{Introduction}

In a theoretical sense, stars are formed in gas and dust clouds with a nonuniform matter distribution and scattered throughout most galaxies. In astrophysics, the term compact object usually  refers collectively to white dwarfs and neutron stars that  form at the end of their stellar evolution. Typically, for such  compact sources it is necessary to investigate the microscopic composition and properties of dense matter on extreme conditions. This is because at such extreme densities nuclear matter may consist not only of nucleons and leptons but also several exotic components in their different forms and phases such as mesons, hyperons and baryon resonances as well as strange quark matter (SQM). However, it is still not possible to find a comprehensive description of the extremely dense matter in a strongly interacting regime. Therefore it is useful to investigate an exact composition and the nature of particle interactions in the interior of this kind of object.  To determine the structure of a compact star within the framework of the general theory of relativity, a widely followed route is to specify an equation of state and then solve the Einstein field equations. Customarily this avenue has proved fruitful when the law of energy conservation is used in the form of the  Tolman-Oppenheimer-Volkoff (TOV) equation
(see \cite{Oppenheimer,Tolman}) or the equation of hydrodynamical equilibrium.

It is possible that anisotropic matter is an important ingredient in many astrophysical objects such as stars, gravastars etc. Historically, considerable effort has been dedicated to gaining a comprehensive understanding of the properties of  anisotropic matter, with the hope of producing physically  viable models of  compact stars.  In particular, compact stars may soon provide information about the gravitational interaction in an extreme gravitational environments. Their extreme internal density and strong gravity hints that pressure within such compact objects may not be in the form of  a perfect fluid i.e., there exist two different kinds of interior pressures namely, the radial and tangential pressure \cite{Herrera4}. Such models may be evolved to study  phase transitions and distributions involving the  mixture of two fluids  \cite{Sawyer}. This effect was first pointed out  by Lema$\hat{i}$te  \cite{Lema}, in the structure and evolution of compact objects. However, interest  in the study of anisotropic relativistic matter distributions in general relativity has been rekindled by  Bowers and Liang \cite{bowers}. They obtained a static spherically symmetric configuration and  analyzed changes in the  surface redshift and gravitational mass by  generalization of the equation of hydrostatic equilibrium.
The theoretical investigations by  Ruderman in \cite{Ruderman} pointed out that at very high densities of order $10^{15}$  g/cm$^3$  nuclear matter tends to become anisotropic in nature. In their work, they consider that for massive stellar objects the radial pressure may not be equal to the tangential one.  At these physical conditions different arguments have been introduced for the existence of anisotropy in star models such as by the presence of type 3A superfluid \cite{Kippenhahm}, different kinds of phase transitions \cite{Sokolov},  the presence of solid core, mixture of two fluids or by other different physical phenomena.

Consequently, in order to understand the peculiar properties of matter in a state of highly anisotropic pressure a large number of works on anisotropic fluid spheres are available in different literature. The particular case of electromagnetic mass model Herrera and Varela \cite{Herrera7}  introduced a condition of anisotropy parameter in the form $P_t-P_r =  gq^2r^2$ where g is a non-zero constant, whereas Barreto \textit{et al.} \cite{Barreto} considered  electrically charged matter as anisotropic matter and so on. On the other hand, Mak and Harko in \cite{Mak} found an exact solution of Einstein's field equations for an anisotropic fluid sphere.  In a recent treatment, Herrera and Barreto studied polytropes for anisotropic matter both in the Newtonian \cite{Herrera} and the general relativistic regimes \cite{Herrera1,Herrera2}. It should be worth noting that  a simple algorithm for all static spherically symmetric anisotropic solutions of Einstein's equations have been analyzed in \cite{Herrera5}. Models for charged anisotropic solutions with a quadratic equation of state have been in \cite{Feroze,Maharaj2}. Maharaj and Maartens \cite{Maharaj1} have critically examined models of static anisotropic fluid spheres under the assumption of uniform energy density.

Characteristically the mathematical problem of developing models of anisotropic fluid spheres amounts to solving a coupled system of three independent nonlinear partial differential equations in five geometrical and dynamical variables namely the metric potentials ($\nu$) and ($\lambda$) and the density ($\rho$), radial pressure ($p_r$) and tangential pressure ($p_t$). Because the system is underdetermined it is possible for any metric to solve the system of field equations. This approach is not necessarily productive as all control over the physics of the problem is relinquished. For example, an equation of state is not likely to be present and this is often viewed as a standard for perfect fluids. Note that the difference between the pressures $p_r - p_t$ is known as the anisotropic parameter denoted by $\Delta$. The approach we follow in this paper is to specify one of the gravitational potentials as the Vaidya-Tikekar \cite{VT} potential which has been shown to model superdense stars. Then we specify the behavior of the anisotropy parameter with the additional help of Lake's potential. Finally we endeavor to solve the pressure anisotropy equation to reveal the general behavior of the remaining gravitational potential. Once the model is complete we are in a position to investigate its physical properties.

In the present paper, our main motivation is to obtain an exact solution for a static anisotropic fluid sphere to the Einstein equations, employing the Korkina and Orlyanskii \cite{Korkina} ansatz for the metric potential. The outline of the paper will be as follows: Following a brief  introduction in Sec. \textbf{I}, we consider a spherical symmetric metric and present the structure equations for anisotropic fluid distributions in Sec. \textbf{II}. Paying particular attention to solving the system of equations analytically, we assume a particular form of  metric potential and obtain the expression for density and pressures in Sec. \textbf{III}. Next, in Sec. \textbf{IV}, we discuss some physical features of the model maintaining  the regularity and matching conditions for the solution and obtained results compared with observational data. Finally, Sec. \textbf{V}, is devoted to closing remarks.

\section{Metric and The Einstein Field Equations}

Consider the metric for  static fluid distributions with spherical symmetry is given by
\begin{equation}
ds^{2} = e^{\nu(r) } \, dt^{2}-e^{\lambda(r)} dr^{2} -r^{2}(d\theta ^{2} +\sin ^{2} \theta \, d\phi ^{2}),
\label{eq1}
\end{equation}
where  $\nu = \nu(r)$ and $\lambda = \lambda(r)$ are the two unknown metric functions
of the radial coordinate $r$ alone. These potentials uniquely determine the surface redshift and gravitational mass functions, respectively. The matter content is assumed to be that of an anisotropic fluid.  Such a stress-energy tensor can be written as

\begin{equation}
\label{eq2}
T_{\mu\nu} = (\rho + p_t)u_\mu\,u_\nu + p_t{g_{\mu\nu}} + (p_r - p_t) \chi^\mu \chi_\nu,
\end{equation}
where $u_\mu$ is the four-velocity and $\chi_\mu$ is the unit spacelike
vector in the radial direction. Thus, the Einstein field equation, $G_{\mu\nu} = 8\pi T_{\mu\nu}$
provides the following gravitational field equations ($G_{\mu\nu}$ is the Einstein tensor)
\begin{equation}
\label{eq3}
\kappa \, \rho(r) = \frac{\lambda '}{r} e^{-\lambda } +\frac{(1-e^{-\lambda } )}{r^{2}},
\end{equation}
\begin{equation}
\label{eq4}
\kappa \, p_{r}(r) = \frac{\nu'}{r} e^{-\lambda } -\frac{(1-e^{-\lambda } )}{r^{2} },
\end{equation}
\begin{equation}
\label{eq5}
\kappa \, p_{t}(r) = \left[\frac{\nu''}{2} -\frac{\lambda' \nu'}{4} +\frac{{\nu'}^{2} }{4} +\frac{\nu'-\lambda '}{2r} \right]\, e^{-\lambda },
\end{equation}
where the prime denotes a derivative with respect to the radial coordinate, r.
Here, $\rho$ is the energy density, while $p_r$ and $p_t$ are radial and
transverse pressures of the fluid distribution. We consider this discussion by $p_t \neq  p_r$.
Consequently, $\Delta$ = $p_t - p_r$ is denoted as the anisotropy factor according to
Herrera and Leon \cite{Herrera3}, and its measure
the pressure anisotropy of the fluid. It is to be noted that at
the origin $\Delta = 0$  is a particular case of an isotropic pressure. Using Eqs. (\ref{eq4}) and (\ref{eq5}), one can obtain the simple form of anisotropic factor
\begin{equation}
\label{eq6}
\Delta =\kappa \, (p_{t} -\, p_{r} ) = e^{-\lambda } \left[\frac{\nu''}{2} -\frac{\lambda' \nu'}{4} +\frac{{\nu'}^{2} }{4} -\frac{\nu'+\lambda '}{2r} -\frac{1}{r^{2} } \right]\,  +\frac{1}{r^{2} },
\end{equation}
However, a force due to the anisotropic nature is represented by $\Delta$/r,
which is repulsive, if $p_{t} > p_{r}$, and attractive if $p_{t} < p_{r}$ of
the stellar model. Throughout the discussion we assume the natural units $\kappa=8\,\pi $  and $G=c=1$.
\begin{figure}[h!]
\centering
\includegraphics[width=7cm]{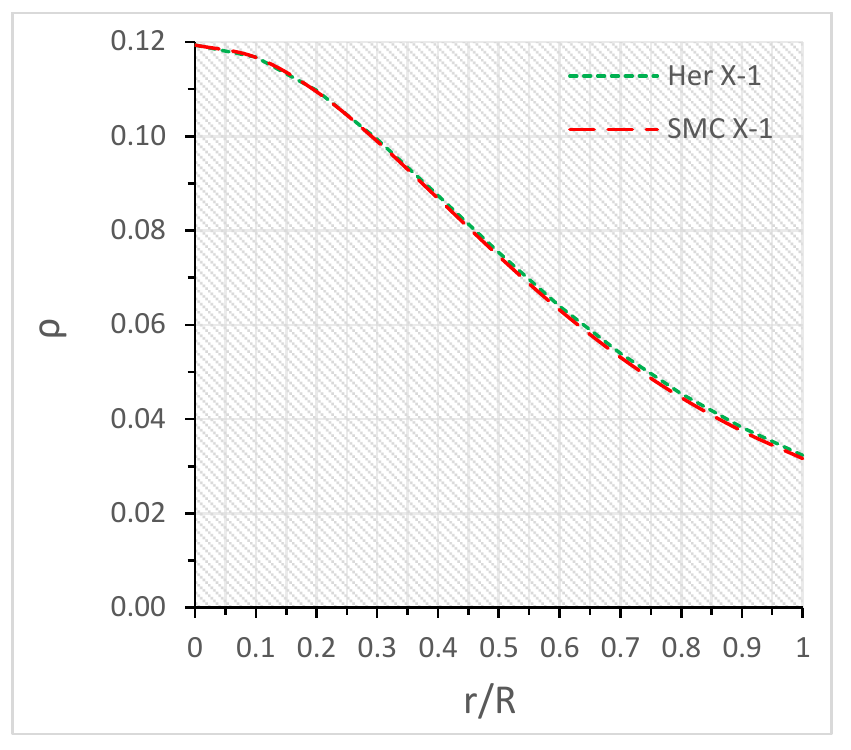} \includegraphics[width=7cm]{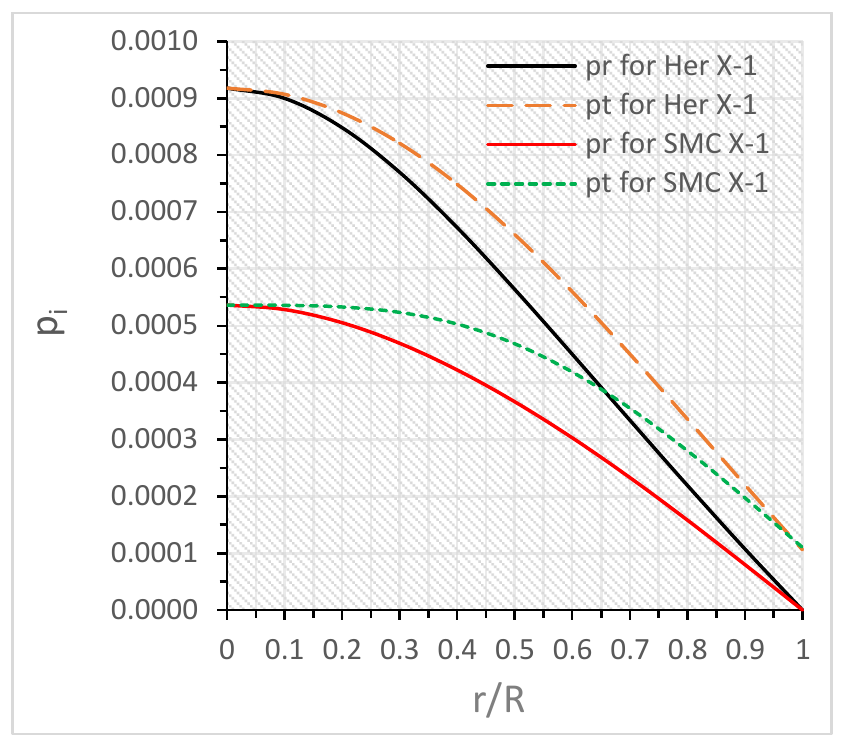}\\
\includegraphics[width=7cm]{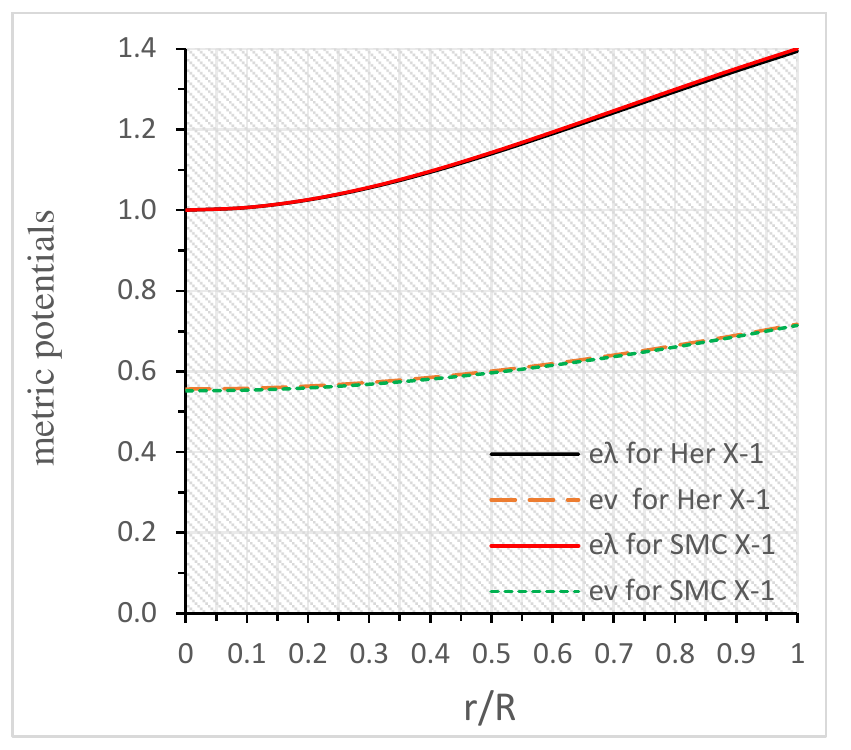} \includegraphics[width=7cm]{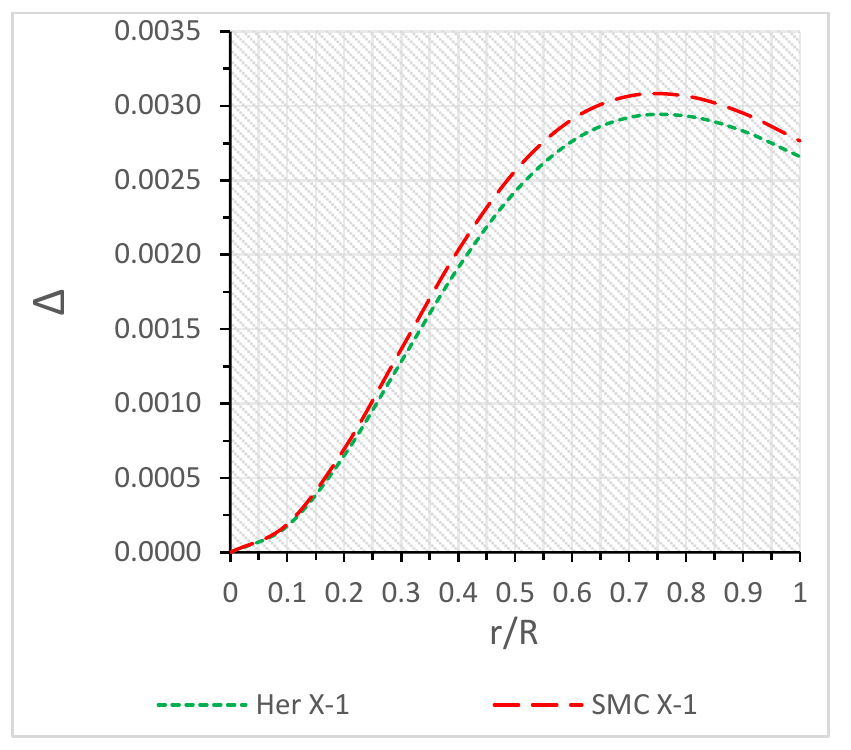}
\caption{\emph{Variation for energy density, radial and transverse pressures have been plotted  against the radial parameter, while the metric potentials and  anisotropic function (Eq. (\ref{eq10}) have been displayed in the second row for the compact star Her X-1 and SMC X-1. The parameter values which we have used given in Table-\,\ref{Table11-1}}}
\label{f1}
\end{figure}

\section{Exact solution of the models for anisotropic stars}

In seeking solutions to Einstein’s field equations for an anisotropic fluid matter we have five unknown functions of $r$, namely, $\rho$(r), $p_r$(r), $p_t$(r), $\nu(r)$ and $\lambda(r)$. In place of the pressure functions, we may invoke the anisotropy parameter $\Delta$ expressing the difference between the tangential and radial pressures.   In either case we have 3 equations and 5 unknown functions. For this reason, to solve these equations analytically one has to specify two variables {\it{ a priori}}. We will demonstrate this feature in an explicit manner for  physically acceptable stellar models.

Introducing  the metric ansatz
\begin{equation}
\label{eq7}
e^{-\lambda(r)} =\frac{1+Cr^{2} }{1+2\,Cr^{2} },
\end{equation}
and the redefinition $y(x)=e^{\nu/2}$ where $x = Cr^2$ for some constant $C$, the field equations may be transformed to an equivalent form conducive to locating exact solutions more efficiently. This form of the metric potential was initially considered by Vaidya and Tikekar \cite{VT} in studying spheroidal spacetimes governing the behavior of superdense stars and subsequently utilized in the work of Korkina and Orlyanskii \cite{Korkina}. The Korkina-Orlyanskii model was  extensively studied in \cite{Gupta}. Observe  that this choice of metric potential yields a singularity free solution at $r = 0$ and the metric coefficient is $e^{\lambda(0)}$ = 1. This will become relevant in the physical analysis later.
Using Eq. (6) and Eq. (7), we obtain
\begin{equation}
\label{eq8}
\frac{d^2y}{dx^2}-\frac{1}{2\,(1+x)\,(1+2x)}\,\frac{dy}{dx}+\frac{(1+2x)^2}{4\,x\,(1+3x+2x^2)}\,
\left[\frac{2\,x}{(1+2x)^2}-\frac{\Delta}{C}\right]\,y=0,
\end{equation}
in our transformed coordinates.
The above Eq.(\ref{eq8}) incorporates  the anisotropy factor $\Delta$. Note that
in the case of an isotropic pressure $\Delta =0$ then at the centre  $y=(1+x )^{1/2}$ and we have a particular solution of  Eq.(\ref{eq7}). This form of $y$ has physical  relevance as discussed by  Lake  \cite{Lake}.

The master equation (\ref{eq8}) is a second order differential equation and is difficult to  solve by standard techniques. Moreover it is still under-determined. We elect to prescribe the anisotropic parameter $\Delta$ with the help of a slightly modified version of Lake's \cite{Lake} potential in the form
\begin{equation}
\label{eq9}
y=(1+x -\beta\,x )^{1/2} , ~~~ \beta>0 ,
\end{equation}
for a positive constant $\beta$.
This approach is not novel (as can be verified in the work in \cite{Maurya11}). Substituting the modified Lake potential for $y$ from  Eq.(9) in Eq.(8) we obtain
\begin{equation}
\label{eq10}
\Delta =\frac{\beta\,C\,x\,[3-\beta-4\,\beta\,x+4\,x]}{(1+2\,x)^2\,(1+x -\beta\,x)^2},
\end{equation}
for the measure of pressure anisotropy.
Obviously  $\Delta=0$ when $\beta=0$ thus regaining  the neutral isotropic perfect fluid solution of solution of Korkina and Orlyanskii [21]. However for a physical anisotropic solution the anisotropy factor $\Delta$ must be finite and positive. Then using the value of $\Delta$ from Eq.(\ref{eq10}) in Eq.(\ref{eq8}), we have
\begin{equation}
\label{eq11}
\frac{d^2y}{dx^2}-\frac{1}{2\,(1+x)\,(1+2x)}\,\frac{dy}{dx}+\frac{(1+2x)^2}{4\,x\,(1+3x+2x^2)}\, \left[\frac{2\,x}{(1+2x)^2}-\frac{\beta\,x\,[3-\beta +4\,(1-\beta)x]}{(1+2\,x)^2\,(1+x -\beta\,x)^2}\right]\,y=0,
\end{equation}
which is a second order linear differential equation in $y$.
Reverting to the original coordinates the exact solution of the above differential equation may be expressed as
\begin{eqnarray}
\label{eq12}
y = (1+Cr^{2} -\beta\,Cr^2)^{1/2}\,\left[A+ B\,\left(\sqrt{2}\,\ln \left|\frac{(1-\beta)\,(4\,Cr^2+3)}{4}+\frac{F(r)}{\sqrt{2}}\right|\,\right)+G(r)\right],
\end{eqnarray}
where we denote
$x = Cr^2$, ~$F(r)=\sqrt{(1-\beta)^2\,(2\,C^2r^4+3\,Cr^2+1)}$, ~~~$G(r)=-\sqrt{\frac{\beta+1}{\beta}}\,\ln \left|1+\frac{2\,\beta (\beta+1)+2\,\sqrt{\beta (\beta+1)}\,F(r)}{(3\beta+1)\,(1+Cr^{2} -\beta\,Cr^2)}\right|$.

The obtained metric function $y = e^{\nu/2}$ is finite and monotonically increasing throughout the stellar interior as depicted in  Fig. 1 (second row, left panel). Accordingly  we obtain the energy density, radial pressure and transverse pressures in the form

\begin{equation}
\label{eq13}
\frac{8\,\pi \,p_r}{C} = \frac{(1-2\,\beta\,-\beta\,Cr^2+2Cr^2)}{(1+2Cr^2)\,(1+Cr^2-\beta\,Cr^2)}+\frac{4(1-\beta)\,B}{y}\,\left[ 2+\frac{(1+\beta)}{(1+Cr^2-\beta\,Cr^2)} \frac{f(r)}{g(r)}-\frac{2\,\sqrt{\beta\,(\beta+1)}\,(\beta+1)\,h(r)}{(1+Cr^2-\beta\,Cr^2)\,g(r)}\right],
\end{equation}
\begin{equation}
\label{eq14}
\frac{8\,\pi \,\rho}{C} =  \frac{3+2Cr^2}{(1+2Cr^2)^2},
\end{equation}
\begin{equation}
\label{eq15}
\frac{8\,\pi \,p_t}{C} =\frac{8\,\pi \,p_r}{C}+\frac{\beta\,C r^2\,[3-\beta-4\,\beta\,Cr^2+4\,Cr^2]}{(1+2\,Cr^2)^2\,(1+Cr^{2} -\beta\,Cr^2)^2},
\end{equation}
where we have set \\

$f(r)=(1-\beta)\,[1+Cr^2+\beta\,(2+3\,Cr^2)]$, ~~

$g(r)=(1-3\,\beta)\,(1+Cr^2-\beta\,Cr^2)+2\,\sqrt{\beta\,(\beta+1)}\,\left[\sqrt{\beta\,(\beta+1)}+h(r)\right]$, \\

$h(r)=\sqrt{(1+\beta)^2\,(2\,C^2r^4+3Cr^2+1)}$ for simplicity.

Observe again from  Eq.(\ref{eq15}) that we have $p_t = p_r$, when $r=0$. This is expected at the center of the star. To further examine the physical character of these solutions for  physical admissibility it is required that

$\bullet$ the energy density is positive definite and its gradient is negative everywhere within the stellar interior.

$\bullet$ for an anisotropic fluid distribution radial and tangential pressures are positive definite and the radial pressure gradient is negative within the radii.

To examine the consequences more closely we take the first order differentiation with respect to the radial coordinate and obtain
\begin{equation}
\label{eq16}
\frac{8\,\pi \,dp_r}{C\,dr} =2\,C\,r\,\left[-\frac{4\,P_1(r)}{(1+2Cr^2)^2}\, +\frac{4\,(1+Cr^2)\,[P_2(r)-P_3(r)-P_4(r)]}{(1+2Cr^2)}\,+\frac{2}{(1+2Cr^2)^2}\right],
\end{equation}
\begin{equation}
\label{eq17}
\frac{8\,\pi \,dp_t}{C\,dr} = \frac{8\,\pi \,dp_r}{C\,dr} +\frac{2\,\beta\,C\,r\,[-3+Cr^2+18C^2r^4+16C^3r^6+\beta^2\,Cr^2\,(1+6\,Cr^2+16\,C^2r^4)-\beta\,P_5(r)]}{(1 + 2\, Cr^2)^3\,~~ (1 + Cr^2 - \beta\,Cr^2)^3},
\end{equation}
and
\begin{equation}
\label{eq18}
\frac{8\,\pi \,d\rho}{C\,dr} = - \frac{4\,C\,r\,(5+2\,Cr^2)}{(1+2\,Cr^2)^3},
\end{equation}
For notational simplicity we choose

$P_1(r)=\left[ \frac{(1-\beta)}{2\,(1+Cr^2-\beta\,Cr^2)} -\frac{B\,(1-\beta)\,(1+2\,Cr^2)}{y\,\sqrt{1+Cr^2-\beta\,Cr^2}\,~\sqrt{1+3\,Cr^2+2\,C^2r^4}}\right]$,

$P_2(r)=\frac{B\,(1-\beta)}{\sqrt{1+Cr^2-\beta\,Cr^2}\,~\sqrt{1+3\,Cr^2+2\,C^2r^4}}  \left[ \frac{(1-\beta)}{(1+Cr^2-\beta\,Cr^2)}-\frac{1}{2\,(1+3\,Cr^2+2\,C^2r^4)} \right]$,

$P_3(r) = \frac{B\,(1-\beta)^2\,(1+2\,Cr^2)}{(1+Cr^2-\beta\,Cr^2)^{3/2}\,(1+3\,Cr^2+2\,C^2r^4)}+\frac{(1-\beta)^2\,y}{(1+Cr^2-\beta\,Cr^2)^{2}}$,

$P_4(r)=\left[ \frac{(1-\beta)}{2\,(1+Cr^2-\beta\,Cr^2)} -\frac{B\,(1-\beta)\,(1+2\,Cr^2)}{y\,\sqrt{1+Cr^2-\beta\,Cr^2}\,~\sqrt{1+3\,Cr^2+2\,C^2r^4}}\right]^2$,

$P_5(r)=\left[-1 - 2 Cr^2 + 24 C^2r^2 + 32 C^3r^6\right]$.\\
which convey information on the  maximum value of the  central density and central pressure. Interestingly, the radial pressure $p_r$ vanish but  the tangential pressure $p_t$ does not vanishes at the boundary (see Fig.\,\ref{f1},).

\section{Physical features and stability of anisotropic compact stars }

To confirm that we are not losing essential physics for a stellar structure at the interior and outer radius, we perform some analytical calculations. Then we discuss how the equilibrium structure and stability of strange stars are affected due to anisotropic pressure. In essence, this is done by studying general physical properties and plotting several figures for some of the compact star candidates. The solutions found in this paper may be used to study relativistic compact stellar objects.

\subsection{Boundary Condition}

It is known that all astrophysical objects are immersed in vacuum or almost vacuum spacetime and at  the juncture interface we match the interior spacetime $(\mathcal{M}_{-})$ to an appropriate exterior vacuum region $(\mathcal{M}_{+})$. In the case at hand, the exterior is described by the Schwarzschild geometry, i.e.,
\begin{equation}
\label{eq19}
ds^{2}=\left(1-\frac{2M}{r}\right)dt^{2}-\frac{dr^{2}}
{1-\frac{2M}{r}}-r^{2}(d\theta^{2}+\sin^{2}\theta d\phi^{2}),
\end{equation}
where $M$ is the mass within a sphere of radius $R$. In order to match smoothly on the boundary surface $r = R$, we impose the Israel-Darmois junction conditions for this system which are tantamount to  the following two conditions \cite{Synge}:

\begin{subequations}
\label{eq20}
\begin{align}
    e^{-\lambda(R)} = 1-\frac{2M}{R}~~~ \text{and}~~~ e^{\nu(R)} = 1-\frac{2M}{R}\label{subeq1},\\
    p_r(r=R) = 0 \label{subeq2},
\end{align}
\end{subequations}
from which  the  constants $A$, $B$ and $C$ may be determined. Furthermore, using the expression (\ref{eq13}), we obtain
\begin{eqnarray}
\label{eq21}
\frac{A}{B} = -\frac{4\,(1-\beta)\,(1+2CR^2)\,\Phi(R)}{(1-2\,\beta+\beta\,CR^2+2\,CR^2)} \left[ 2+\frac{(1+\beta)}{\Phi^2(R)} \frac{F(R)}{G(R)}+\frac{2\,\sqrt{\beta\,(\beta+1)}\,(\beta+1)\,H(R)}{\Phi^2(R)\,G(R)}\right]-\Psi(R),
\end{eqnarray}
and using the condition (\ref{subeq1}), we obtain
\begin{equation}
\label{eq22}
B=\sqrt{\frac{(1+CR^{2} )}{(1+2\,CR^{2})\,(1+CR^{2} -\beta\,CR^2)}}~~\times~ \frac{1}{\left[\frac{A}{B}+ \left(\sqrt{2}\,\ln \left|\frac{(1-\beta)\,(4\,CR^2+3)}{4}+\frac{\Psi_1(R)}{\sqrt{2}}\right|\,\right)+\Psi_2(R)\right]},
\end{equation}

for notational simplicity we introduce\\

$\Phi(R)=(1+CR^2-\beta\,CR^2)^{1/2}$,    $F(R)=(1-\beta)\,[1+CR^2+\beta\,(2+3\,CR^2)]$,\\~~

$G(R)=-(3\,\beta+1)\,(1+CR^2-\beta\,CR^2)+2\,\sqrt{\beta\,(\beta+1)}\,\left[\sqrt{\beta\,(\beta+1)}+H(R)\right]$, \\

$H(R)=\sqrt{(\beta-1)^2\,(2\,C^2R^4+3CR^2+1)}$  , ~~~ $\Psi(R)=\left[\left(\sqrt{2}\,\ln \left|\frac{(1-\beta)\,(4\,Cr^2+3)}{4}+\frac{\Psi_1(R)}{\sqrt{2}}\right|\,\right)+\Psi_2(R)\right]$, \\

$\Psi_1(R)=\sqrt{(1-\beta)^2\,(2\,C^2r^4+3\,Cr^2+1)}$,  ~~~ $\Psi_2(R)=-\sqrt{\frac{\beta+1}{\beta}}\,\ln \left|1-\frac{2\,\beta\, (\beta+1)+2\,\sqrt{\beta \,(\beta+1)}\,\Psi_1(R)}{(3\,\beta+1)\,(1+Cr^{2} -\beta\,Cr^2)}\right|$.

For a given radius $R$, one can determine the total mass $M$ of the star and vice-versa.  It may be mentioned here that bounds on stellar structures, including the  mass-radius ratio as proposed by the Buchdahl-Bondi inequality \cite{Buchdahl,Bondi}, exists and is given by  $\frac{2M}{R} \leq \frac{8}{9}$. This serves as an upper bound on the total compactness of a static spherically symmetric isotropic fluid sphere (the geometric units $c = G = 1$ have been used). In fact, this bound has been updated in the presence of charged gravitational fields \cite{Bohmer1,andr} and for a non-zero cosmological constant  \cite{Mak,Bohmer}. The impact of the mass-radius ratio on the equation of state has been considered by Carvalho {\it{et al}} \cite{carv} for white dwarfs and neutron stars, by Swift {\it {et al}} for exoplanets and for the nuclear centre by Lattimer \cite{lattim}.

\begin{table}
\centering \caption{The approximate values of the masses $M$, radii $R$,
and the constants $A, B$ and $C$ for the compact stars} \label{Table11-1}
\scalebox{0.8}{%
\begin{tabular}{c|c|c|c|c|c|c|c|c} \hline
 \quad Compact Stars \quad\quad\quad & \quad \quad $CR^2$ \quad \quad &\quad $\beta$ \quad\quad   &\quad A \quad\quad & \quad B \quad\quad & \quad C (\text{Km.$^{-2}$}) \quad & \quad M$\left(M_\odot\right)$ \quad & \quad R \quad & \quad M/R \quad \\

\hline Her X-1     & 0.65087  & 0.74391 & 0.2846647  &  -0.11649 & 8.276327$\times10^{-3}$ &  0.8505 & 8.86805 & 0.14146\\

\hline SMC X-1 &  0.66785 & 0.73157 & 0.282161   & -0.114127 & 5.800009$\times10^{-3}$ &   1.04  & 10.7306 & 0.142956\\ \hline
\end{tabular}}
\end{table}

Here, we demonstrate that for some particular values of the parameters and by plugging in the true values for  $c$ and $G$ at appropriate places as given in Table 1, we can generate specific masses and radii of some well known pulsars given by Gangopadhyay {\it{et al}}  (2013), for the objects
Her X-1 and SMC X-1. Some possibilities of such types are tabulated in Table-\,\ref{Table11-1}.

Let us start  by considering the surface gravitational redshift $z_S$ of this compact objects with help of the definition  $z_S$ = $\Delta \lambda/\lambda_{e}$ = $\frac{\lambda_{0}-\lambda_{e}}{\lambda_{e}}$, where $\lambda_{e}$ is the emitted wavelength at the surface of a nonrotating star and the
observed wavelength $\lambda_{0}$ received at radial coordinate r. Thus, the gravitational redshift, $z_S$ from the surface of the star as measured by a distant observer ($g_{tt}(r) \rightarrow - 1$), is given by
\begin{eqnarray}
\label{eq23}
z_S = -1+\arrowvert g_{tt}(r) \arrowvert ^{-1/2} = -1+\left(1-\frac{2M}{R}\right)^{-1/2},
\end{eqnarray}
where $g_{tt}(r)$ = $e^{\nu(R)}$ =$\left(1-\frac{2M}{R}\right)$ is the metric function \cite{zel}. According to Buchdahl \cite{Buchdahl} and Straumann \cite{Straumann} a constraint  on the  gravitational redshift for  perfect fluid spheres is given by  $z_{S} <$ 2 for isotropic stars. However, we may have a situation for an anisotropic star  to admit higher redshifts such as  $z_{S}$ = 3.84, as given in Ref. \cite{Ivanov}. We have summarized our results in Table-\,\ref{Table11-2}, for the stellar structures Her X-1 and SMC X-1 by taking the same values of the constant as mentioned in Table-\,\ref{Table11-1}.

\subsection{Energy conditions}

It is reasonable to expect that models of anisotropic fluids satisfy the energy conditions within the framework of general relativity. There often exists a  linear relationship between energy density and pressure of the matter  obeying certain restrictions.  In view of the above situation, we examine (i) the Null energy condition (NEC), (ii) Weak energy condition (WEC) and (iii) Strong energy condition (SEC) to enhance our investigation of  the structure of relativistic spacetimes. More precisely, we have the following proposition:
\begin{subequations}
\label{eq24}
\begin{align}
\textbf{NEC:}~ \rho(r)+p_r \geq  0,\label{subeq3}\\
\textbf{WEC}_r:~ \rho+p_r \geq  0, ~~\mathrm{and}~~\rho(r) \geq  0, \label{subeq4}\\
\textbf{WEC}_t:~ \rho+p_t \geq 0  ~~\mathrm{and}~~\rho(r) \geq  0,\label{subeq5}\\
\textbf{SEC:}~ \rho+p_r+2\,p_t \geq  0.\label{subeq6}
\end{align}
\end{subequations}

\begin{figure}[h]
\centering
\includegraphics[width=7cm]{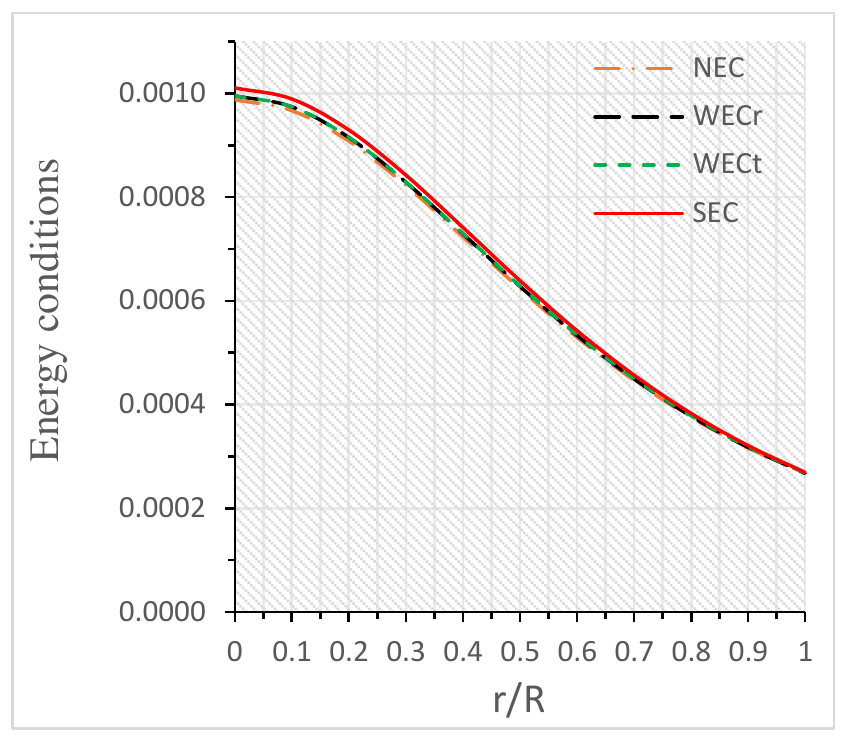} \includegraphics[width=7cm]{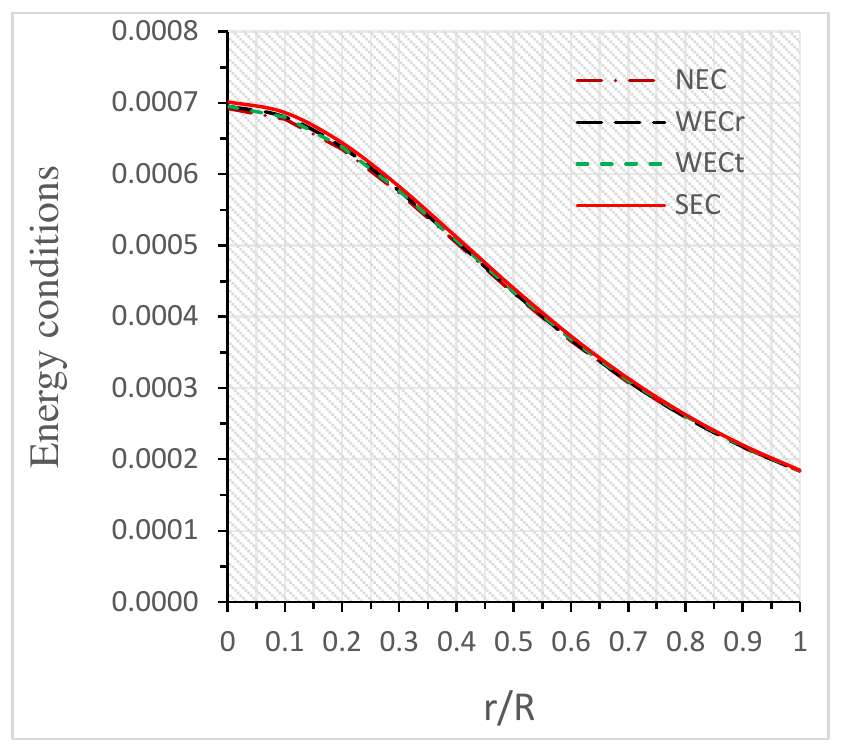}
\caption{\emph{The energy conditions, namely, null energy condition (NEC), weak energy condition (WEC), strong energy condition (SEC) are shown in this figures, for the compact stars SMC X-1 and Her X-1. The numerical values of the constants are given in Table-\,\ref{Table11-1}.}}\label{f2}
\end{figure}
Using the above expression for all the terms in this inequality, one can easily
justify the nature of energy condition for the specific stellar configuration Her X-1 and SMC X-1.
To further interpret these results we use graphical representation of the energy
Conditions, as can be seen from the Fig. 2. For the complicated expression given in equations ((\ref{eq24}a-\ref{eq24}d)), we only write down the
inequalities and plotted the graphs as a function of the radius.
As a result, this is eminently clear from Fig.\,\ref{f2}, that all the energy conditions are satisfied for our proposed model.

\subsection{Generalized Tolman-Oppenheimer-Volkov Equation}

In order to investigate the hydrostatic equilibrium under different forces of compact star  for a physically acceptable model we investigate the gravitational and other fluid forces. By considering the generalized Tolman-Oppenheimer-Volkoff (TOV) \cite{Oppenheimer1939,Leon1993}, equation one can clarify the situation for an anisotropic fluid distribution, which is
\begin{eqnarray}
& \qquad\hspace{-1cm}-\frac{M_G(r)(\rho+p_r)}{r^2}e^{\frac{\lambda-\nu}{2}}-\frac{dp_r}{dr}+\frac{2}{r}(p_t-p_r)=0, \label{eq25}
\end{eqnarray}
where the effective gravitational mass $M_G(r)$ is defined by
\begin{eqnarray}
&\qquad\hspace{1cm} M_G(r)=\frac{1}{2}{{r}^{2}}e^{\frac{\nu-\lambda}{2}}\nu'. \label{eq26}
\end{eqnarray}

Then Eq. (\ref{eq25}) may be simply obtained as
\begin{eqnarray}
&\qquad\hspace{-1cm}-\frac{\nu'}{2}(\rho+p_r)-\frac{dp_r}{dr}+\frac{2}{r}(p_t-p_r)=0, \label{eq27}
\end{eqnarray}

\begin{figure}[h]
\centering
\includegraphics[width=7cm]{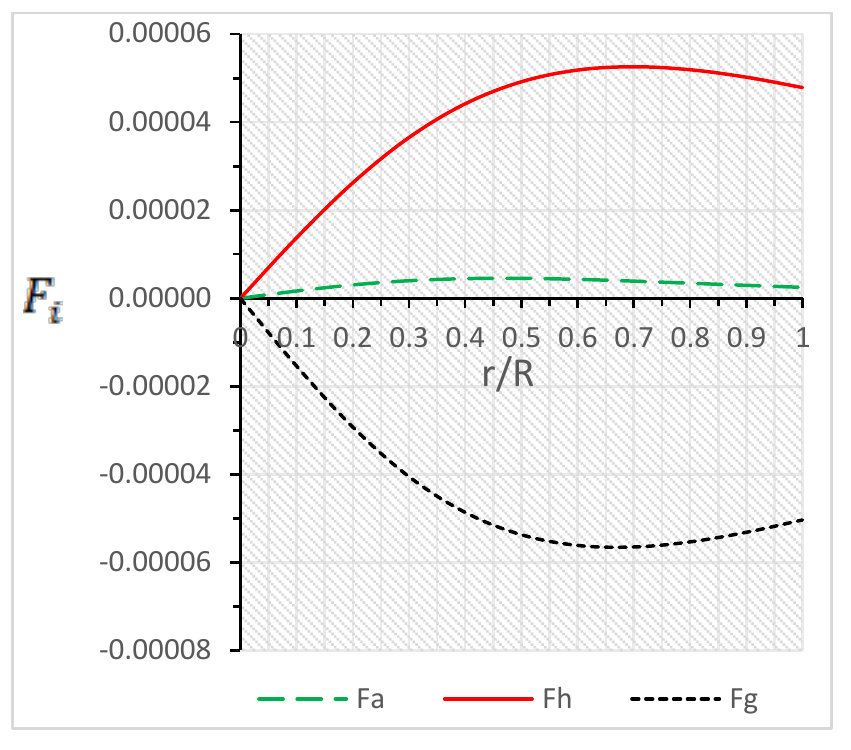} \includegraphics[width=7cm]{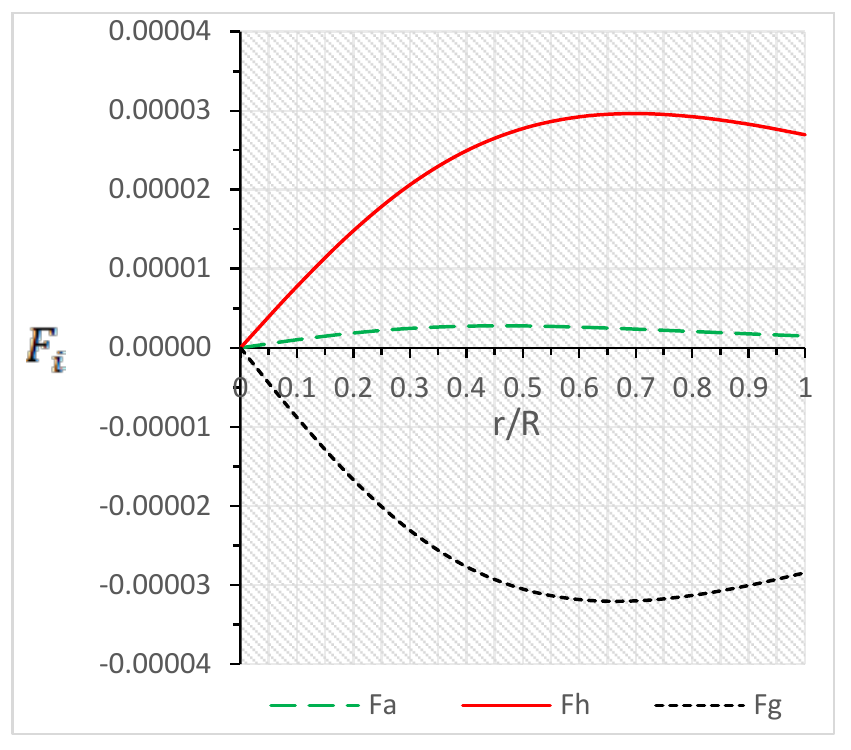}
\caption{\emph{We have plotted different forces, namely, gravitational force ($\mathrm{F_g}$), hydrostatics force ($\mathrm{F_h}$) and anisotropic force ($\mathrm{F_a}$) for studying the effect of the anisotropy in the stability of compact stars. See the subsection C,  for details about  the stable configuration mode.}}
\label{f3}
\end{figure}

In other words, for this case, the TOV equation (\ref{eq27}) expresses   the equilibrium condition for anisotropic fluid spheres subject to gravitational, hydrostatic  plus another force due to the anisotropic pressure. Combined with the above expressions we can write
\begin{equation}
F_g+F_h+F_a=0. \label{eq28}
\end{equation}
Let us now attempt to explain the Eq. (\ref{eq28}) from an equilibrium point of view,
where three different forces are the gravitational force ($F_g$), hydrostatics force ($F_h$) and
anisotropic force ($F_a$) with the expressions:
\begin{eqnarray}
&\quad\hspace{-1.9cm} F_g=  \frac{-\nu'(\rho+p_r)}{2}
= \left[\frac{2\,Cr\,(1-\beta)}{2\,(1+Cr^2-\beta\,Cr^2)}+\frac{2\,C\,B\,r\,\sqrt{1+2\,Cr^2}}{y\,\sqrt{1+Cr^2-\beta\,Cr^2}\,\sqrt{1+Cr^2}}\right]\,(\rho+p_r),\label{eq29}
\end{eqnarray}

\begin{eqnarray}
&\quad\hspace{-0.4cm} F_h=-\frac{dp_r}{dr}
=-\frac{C^2\,r}{4\,\pi}\left[-\frac{4\,P_1(r)}{(1+2Cr^2)^2}\, +\frac{4\,(1+Cr^2)\,[P_2(r)-P_3(r)-P_4(r)]}{(1+2Cr^2)}\,+\frac{2}{(1+2Cr^2)^2}\right], \label{eq30}
\end{eqnarray}

\begin{eqnarray}
 &\quad\hspace{-0.5cm} F_a=\frac{2}{r}(p_t-p_r)
= \frac{2\,\beta\,C^2r\,[3-\beta-4\,\beta\,Cr^2+4\,Cr^2]}{(1+2\,Cr^2)^2\,(1+Cr^{2} -\beta\,Cr^2)^2}. \label{eq31}
\end{eqnarray}
To simplify the above equations we also draw two figures for the compact star candidates Her X-1 and SMC X-1 (  Fig.\,\ref{f3}). Therefore, the claim is that the gravitational force ($F_g$) dominates the hydrostatic ($F_h$) and  anisotropic ($F_a$) forces to maintain the equilibrium condition. In other words, the static equilibrium is attainable due to pressure anisotropy, gravitational and hydrostatic forces, which is evident from Fig.\,\ref{f3}.

\subsection{Stability Analysis}
A very important point that has to be analyzed is the speed of sound propagation $v^2_s $, which is given by the expression $v^2_s = dp/ d\rho$. Naturally the velocity of sound  does not exceed the velocity of light. Thus, the behavior of the sound speed is always less than unity, as we fix here $c = 1$. To analyze the situation we  investigate the speed of sound  along  a radial as well as transverse direction. For an anisotropic fluid distribution and for a stable equilibrium configuration this should  always satisfy $0 < v_{r}^{2}=\frac{dp_r}{d\rho} < 1$ and $0 < v_{t}^{2}=\frac{dp_t}{d\rho} < 1$, as in ref. \cite{Herrera(2016)} for a subluminal sound speed. In looking for charged solutions, Canuto in \cite{Canuto} argued that the speed of sound should decrease  monotonically towards the surface of the star for an ultra-high distribution of matter.

\begin{figure}[h]
\centering
\includegraphics[width=7cm]{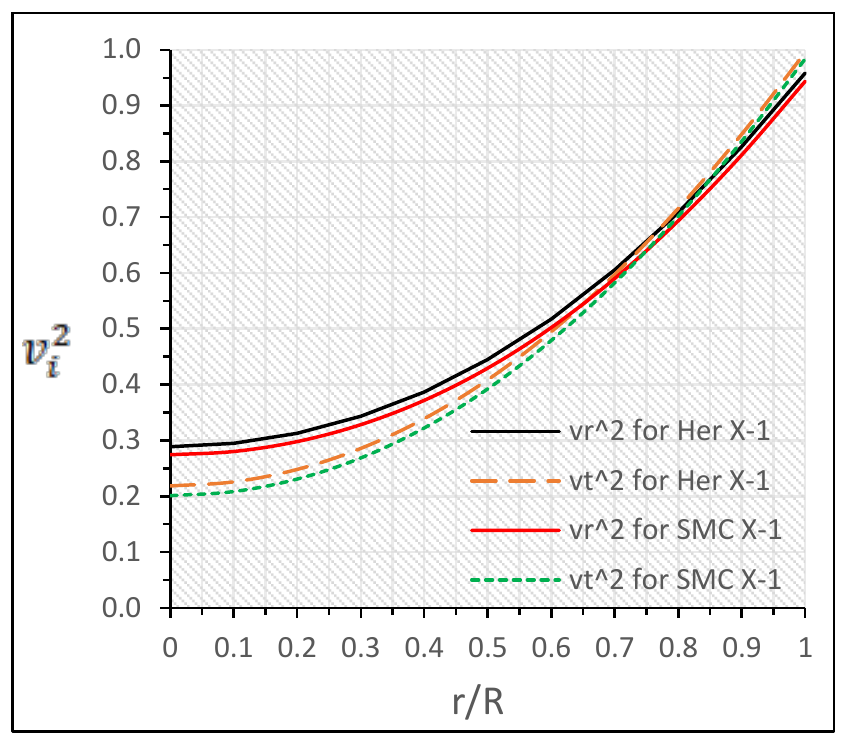}  \includegraphics[width=7cm]{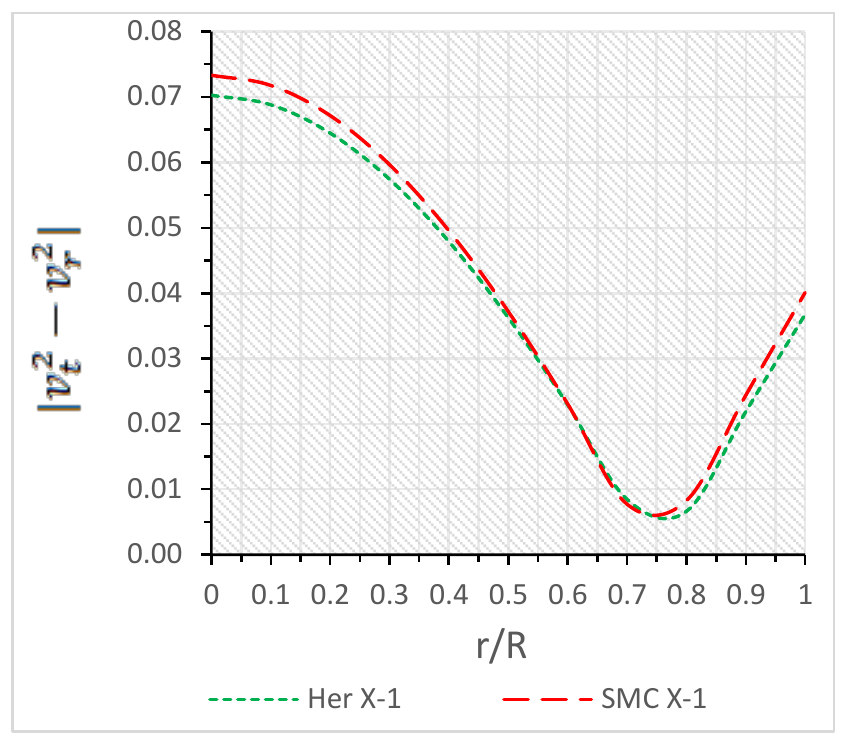}
\caption{\emph{A plot of the characteristic for sound propagation $v_s^2 =d p/d\rho$ along the radial and transverse direction for the same stellar model. We have shown that the speed of sound is less than unity within the stellar radii, by making a choice similar to Fig.\,\ref{f1}} }
\label{f4}
\end{figure}

In our case, the sound velocity has been studied with a graphical representation for an anisotropic fluid distribution. To see this we have plotted figures 4,  for strange star candidates SMC X-1 and Her X-1. As the resulting expressions are very complicated, we illustrate the causality conditions without mathematical explanation and the values of parameters are tabulated in Table- 1. Turning to the case, we see that both $0 \leq  v^2 _{sr} \leq  1$ and $ 0 \leq  v^2 _{st} \leq 1$  everywhere within the anisotropic fluid and monotonic increasing function, which is evident for other compact objects. In addition, it is important to mention the stability of local anisotropic matter distribution, using the concept of Herrera in \cite{Herrera(2016)}. According to this $0< \Bigl\rvert v_{t}^{2}-v_{r}^{2}\Bigl\rvert \leq 1$, for stable potential. In our case, Fig.\,\ref{f4} (right panel) indicates that there is no sign change for the term $v^2_{t}- v^2_{r}$ within the stellar interior.  Therefore, we conclude that our chosen stellar model is stable for our  choice of parameters.


\begin{table}
\centering \caption{\text{The central density, surface density and central
pressure for compact star candidates}.} \label{Table11-2}
\scalebox{0.8}{%
\begin{tabular}{c|c|c|c|c} \hline

   \quad Compact \quad\quad\quad &  \quad Central Density \quad\quad   &\quad Surface Density \quad\quad & \quad Central Pressure \quad\quad & \quad Surface Redshift\quad  \\

 Stars &  $\mathrm{(gm/cm^{3})} $   &  $\mathrm{(gm/cm^{3})}$   & $\mathrm{(dyne/cm^{2})}$ & $\mathrm{Z_S}$ \\ \hline

Her X-1 &  $ 1.333316\times{{10}^{15}}$ & $3.60864\times{{10}^{14}}$ & $9.2297127\times{{10}^{35}}$ & $ 0.180787$ \\ \hline

SMC X-1 & $9.343815\times{{10}^{14}}$ & $2.4753015\times{{10}^{14}}$ & $3.77957\times{{10}^{33}}$ & $ 0.183396$ \\ \hline

\end{tabular}}
\end{table}

\section{Discussion}
In this paper, we have investigated the nature of anisotropic compact stars. Beginning with the Korkina and Orlyanskii \cite{Korkina} ansatz that $e^{-\lambda(r)}$ = $(1+x)/(1+2x)$ where $x = Cr^2$ together with the choice of an anisotropy function $\Delta$ inspired by a prescription of Lake \cite{Lake}, it has been shown that a number of compact objects are compatible with observational data and we have cited Her X-1 and SMC X-1 as  specific examples of this kind of star. Next, by employing the chosen metric functions we simplify the Einstein field equations and study the structure of compact stars.

 Based on physical requirements, we matched the interior solution to an exterior vacuum Schwarzschild spacetime and fixed the constants $A, B,$ and $C$ (see Table-\,\ref{Table11-1} for more details). Then using the values of constant parameters  it is also possible to determine masses and radii for compact stars.  To refine the model further, we show that energy density and pressures are finite at the center and monotonically decreasing towards the boundary which is illustrated in Fig.\,\ref{f1}. For an isotropic compact spheres Buchdahl has provided a bound for spherical mas distributions satisfying the inequality $R>$ (9/8) $R_S$ =  (9/4)G M/$c^2$ \cite{Buchdahl1959} which is stricter than the Schwarzschild bound. From  Fig.\,\ref{f1} (first row extreme right), it can be seen that radial pressure vanishes at the boundary, whilst the tangential one is non-vanishing at the stellar surface.

  In order to investigate the relevance of our model we consider the masses and radii for some well known pulsars Her X-1, and SMC X-1 given by Gangopadhyay \emph{et al.} \cite{Gangopadhyay}  to fit into the observational data. We demonstrate this by using suitable choices of the constant parameters $A, B,$ and $C$ in Table-\,\ref{Table11-1}. In the Table-\,\ref{Table11-2}, we have displayed the surface density of the star Her X-1 and SMC X −1 as 3.60864$\times{{10}^{14}}$ \& $2.4753015\times{{10}^{14}}$ gm/$cm^3$, which is very high and consistent with ultra compact stars \cite{Ruderman,Glendenning,Herzog}. We also note that the gravitational redshift satisfied $z_s \leq 2$ i.e. it is bounded from above as shown in Table-\,\ref{Table11-2}.

  As a future work, we  envisage  investigating  other forms of metric potentials could exhibit more general behaviour and thereby describe other types of compact objects such as  superdense stars for which there exists ample observational data.
\\

\textbf{Acknowledgments}: The author S. K. Maurya acknowledges authority of University of
Nizwa for their continuous support and encouragement to carry out this research work.  AB is thankful to the authority of
Inter-University Centre for Astronomy and Astrophysics, Pune,
India for providing research facilities.

\section*{Appendix: Solution
generating technique for anisotropic fluids: Details }

Since the back drop of such varied developments, Herrera \textit{et al.} \cite{Herrera5} provided a solution generating technique to construct all possible types of solutions of the Einstein field equations for a static spherically symmetric locally anisotropic fluids source  in terms of two generating functions. To see this, let us analyze by using the Eq. (\ref{eq4}) and Eq. (\ref{eq5}), one finds
\begin{equation}
8\,\pi \, (p_{r} -\, p_{t} ) = e^{-\lambda } \left[-\frac{\nu''}{2} +\frac{\lambda' \nu'}{4} -\frac{{\nu'}^{2} }{4} +\frac{\nu'+\lambda '}{2r} +\frac{1}{r^{2} } \right]\,  -\frac{1}{r^{2} }, \label{eq32}
\end{equation}
To facilitate computations, we introduce new variables as
\begin{equation}
e^{\nu(r)}=e^{\left( \int{2\,\Psi(r)-\frac{2}{r}}\right) dr} ~~~\textmd{and}~~~e^{-\lambda(r)}=Z(r) \label{eq33}
\end{equation}
Putting  Eqs.(\ref{eq33}) into the Eq.(\ref{eq32}), yields
\begin{equation}
Z^{\prime}+Z\,\left[\frac{2\\Z^{\prime}}{Z}+2\,Z-\frac{6}{r}+\frac{4}{r^2\,Z}\right]=-\frac{2}{Z}\,\left[\frac{1}{r^2}+\Pi(r)\right], \label{eq34}
\end{equation}
where $\Pi(r) = 8\,\pi\,(p_r-p_t)$.
Integrating Eq.(\ref{eq34}) we obtain $e^{\lambda}$

\begin{equation}
 e^{\lambda(r)}=\frac{\Psi^2\,e^{\int{\left[\frac{4+2\,r^2\,\Psi^2(r)}{r^2\,\Psi(r)}\right]dr}}}
 {r^6\left[-2\,\int{\frac{\Psi(r)\,\left[1+\Pi(r)\,r^2\right]\,e^{\int{\left[\frac{4+2\,r^2\,\Phi^2(r)}{r^2\,\Psi(r)}\right]dr}}}{r^8}\,dr}+D\right]}, \label{eq35}
\end{equation}
which is obtained be Herrera \textit{et al.} \cite{Herrera5}. According to them the the two generating functions are $\Psi(r)$ and  $\Pi$.  Thus, in our case when we introduce $y(r) =e^{\nu/2}$, the above Eq.(\ref{eq33})  turns out to be $\Psi(r)=\left[\frac{y^{\prime}(r)}{y}+\frac{1}{r}\right]$.

As a consequence of the following algorithm, the generating functions in the present case for anisotropic fluid distribution as follows (using the Eq.(\ref{eq10}) and Eq.(\ref{eq12})):\\
\begin{equation}
\Psi(r)=\frac{C\,r\,(1-\beta)}{(1+Cr^2-\beta\,Cr^2)}
+\frac{2\,\widetilde{B}\,C\,r\,\sqrt{1+2\,Cr^2}}{y\,\sqrt{1+Cr^2}\,\sqrt{1-\beta\,Cr^2+Cr^2}}+\frac{1}{r},
\end{equation}

\begin{equation}
\Pi=-\frac{8\,\pi\,\beta\,C^2r^2\,[3-\beta-4\,\beta\,Cr^2+4\,Cr^2]}{(1+2\,Cr^2)^2\,(1+Cr^{2} -\beta\,Cr^2)^2},
\end{equation}
where $\beta$ is positive constant as indicated in Eq.(\ref{eq9}) with $\widetilde{B}$ as a constant of integration. \\

{ }

\end{document}